# Characterization and Thermal Management of a DC Motor-Driven Resonant Actuator for Miniature Mobile Robots with Oscillating Limbs

D. Colmenares[1], R. Kania[1], M. Liu[1], and M. Sitti[2]

*Abstract* - **In this paper, we characterize the performance of and develop thermal management solutions for a DC motor-driven resonant actuator developed for flapping wing micro air vehicles. The actuator, a DC micro-gearmotor connected in parallel with a torsional spring, drives reciprocal wing motion. Compared to the gearmotor alone, this design increased torque and power density by 161.1% and 666.8%, respectively, while decreasing the drawn current by 25.8%. Characterization of the actuator, isolated from nonlinear aerodynamic loading, results in standard metrics directly comparable to other actuators. The micro-motor, selected for low weight considerations, operates at high power for limited duration due to thermal effects. To predict system performance, a lumped parameter thermal circuit model was developed. Critical model parameters for this micro-motor, two orders of magnitude smaller than those previously characterized, were identified experimentally. This included the effects of variable winding resistance, bushing friction, speed-dependent forced convection, and the addition of a heatsink. The model was then used to determine a safe operation envelope for the vehicle and to design a weight-optimal heatsink. This actuator design and thermal modeling approach could be applied more generally to improve the performance of any miniature mobile robot or device with motor-driven oscillating limbs or loads.**

*Index terms* - **resonant actuator, DC motor, thermal model, flapping wing, micro air vehicle**

## I.  INTRODUCTION

DC motors are commonly used to power miniature robots capable of running [1-3], climbing [4], and flying [5-8]. Such behaviors with periodic limb motion require significant torque to continuously accelerate and decelerate the limb. Transmissions have been used to convert rotary motor output into periodic limb motion and often use gearing to increase torque, which declines sharply with decreasing motor size. However, they can be complex and heavy [6], limit system energy efficiency [9], and complicate control [10]. In order to build smaller, more powerful robots, we utilize a resonant actuator inspired by biological muscles, which acts as an elastic actuator, storing energy to reduce inertial power needed to produce these behaviors. For a spring-mass-damper system with periodic actuation occurring at the resonant frequency, the elastic element stores the energy necessary to accelerate the load. Therefore, when operating at resonance, the actuator only needs to provide torque to overcome damping. For flapping, the torque needed to accelerate the wing is almost seven times larger than the torque needed to overcome aerodynamic damping [11]. Therefore, resonant actuation significantly decreases the power requirements of a flapping wing micro air vehicle (FWMAV). Furthermore, resonant actuators can directly drive limbs to efficiently generate high force output with independent control. This is particularly useful for weight-constrained systems where maximizing actuator power density is critical for operation.

The resonant actuator is based on work by Campolo *et al.* [11, 12], where a torsional spring was connected in parallel with the shaft of a brushed DC motor. Our previous work demonstrated the use of a resonant actuator based on the GM15 gearmotor in the FWMAV shown in Figure 1, capable of hover, with a flapping frequency of 20 Hz, weight of 3.2 g, and peak lift to weight ratio of 3.7 [13, 14]. Addition of elastic energy storage to flapping transmission mechanisms has been tested in work by Lau *et al.* as well as Beak *et al.*, although these systems have not been capable of liftoff [9, 15]. Recent work by Roll *et al.* demonstrates liftoff of a vehicle using a custom electromagnetic actuator that achieves resonance with a virtual magnetic spring [16]. However, analysis of the actuator isolated from nonlinear aerodynamic loading has never been performed. Such analysis is critical for improved modeling of the actuator and comparison to systems in the literature. Furthermore, it allows researchers to determine if their work could benefit from this actuation technique, extending its use to other applications. Despite the benefits of resonant actuation, the capacity to dissipate thermal losses fundamentally limits motor performance. Therefore, maximizing output results in a thermal management problem.

Although motors are commonly used in robotic systems, thermal considerations are often neglected when they are operated continuously within the motor specifications.

This work is supported by the National Science Foundation Graduate Research Fellowship Program under Grant No. DGE-1252522.

[1]David Colmenares, Randall Kania, and Miao Liu are with the Department of Mechanical Engineering, Carnegie Mellon University, Pittsburgh, PA 15213, USA, [dcolmena, rkania, miaol1]@andrew.cmu.edu.
[2]Metin Sitti is a director of Max Planck Institute for Intelligent Systems, Stuttgart 70569, Germany and professor at Carnegie Mellon University, Pittsburgh, PA 15213, USA, sitti@is.mpg.de.

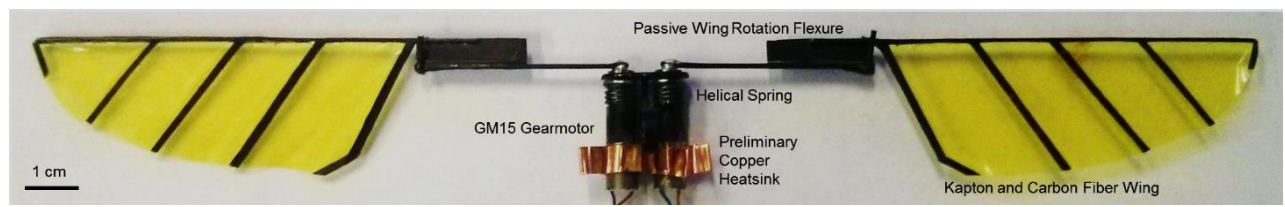

Figure 1. Photograph of the two motor flapping wing prototype with initial copper fin design for thermal management.

However, thermal behavior is critical when intermittent, high-power operation is desired. While early studies focused on continuous operation of industrial-sized devices [17-19], recent work has addressed intermittent operation for motors at the 100 gram scale [20, 21]. The main source of thermal energy is Joule heating of the motor windings produced at a rate of $I^2R$, where $I$ is the input current and $R$ is the winding resistance. In large systems this is the dominant heat source and heat flow is limited by low air-gap convection due to internal laminar flow [22]. However, heat transfer through the motor varies substantially across designs and size scales [23].

The motor in this study weighs 1 gram, two orders of magnitude smaller than those previously characterized, and has been used in several miniature robotic systems [7, 24-26]. A second-order lumped parameter approach was used for the thermal model [27, 28]. Experimental model parameter fitting identified effects of variable winding resistance and bushing friction. Furthermore, high rotor velocities were found to substantially increased air-gap convection. The addition of a heatsink, which increases the surface area to volume ratio of the motor, was also included. While previous studies used temperature measurements from internal components, the following procedure uses only case temperature for fitting. The developed model is validated for dynamic operation of the FWMAV, used to predict safe operating limits, and to optimize a heatsink. This new small scale model and training approach can be applied to accurately characterize micro-motors used in miniature robotic systems.

Following, details are provided for the resonant actuator characterization and key performance metrics are discussed in Section II. The metric of power density is discussed in the context of the literature in Section III. The thermal model and important parameters are described in section IV. The experimental setup, training procedure, and resulting model improvements are discussed in Section V. Section VI presents the trained model output with experimental validation. A conclusion on modeling and improving performance of micro-motor systems is given in Section VII.

## II. DYNAMIC ACTUATOR CHARACTERIZATION

The resonant actuator and gearmotor were characterized using a proof mass, a purely inertial load, in order to isolate the system from the complex aerodynamic loading generated by the flapping wing. The mass is chosen to have a similar moment of inertia to the FWMAV wings in [13]. They were driven bi-directionally with a sinusoidal voltage to produce periodic motion with a constant amplitude. First, the driving frequency was swept from 10 to 30 Hz to identify the resonant frequency where load amplitude was maximized. Both systems were then characterized at 20 Hz with successive tests at increasing voltage amplitudes. The frequency range was set by the torsional spring stiffness, an average of 18 N·mm/rad over a 180° displacement. The spring was fabricated from 1080 steel wire and 87%-118% lighter than equivalent commercial springs [29]. This stiffness can be varied, although stiffening the spring would increase loading on the gearbox and could lengthen the startup transient, or the number of cycles before maximum amplitude is achieved.

Input power was calculated as the product of root-mean-square (RMS) current, measured using a current sensor (ASC712-30A), and RMS voltage, measured across the motor with a DAQ board (National Instruments PCIe-6353), at a resolution of 40 µW. The electrical characteristics of the winding were considered purely resistive as inductance was negligible. Power factor was approximately one as the current and voltage signals were seen to be in phase. Proof mass kinematics were measured from high-speed video (PCO Dimax). Mechanical power was calculated using inverse dynamics analysis as follows: angular velocity and acceleration are derivatives of load position, effective speed is the product of average angular velocity and frequency, torque is the product of load inertia and angular acceleration, power is the product of torque and angular velocity. Maximal performance is shown in Table 1. Calculations were based on 76 steady state cycles.

The AC steady-state gearmotor data closely matches the DC ratings, although maximum torque is achieved with high amplitude oscillations instead of at stall. However, load position was not consistent and increasing the voltage above 4.5 V did not increase load amplitude. The resonant actuator produced 193.8% more torque than the gearmotor and achieved consistent and symmetric high amplitude oscillations. The resonant actuator achieved over 87% efficiency for peak-to-peak oscillation amplitudes above 32.75°, indicating that the spring effectively stores the energy to overcome inertial loading. Furthermore, the mechanical and electrical impedances of the load and motor were well matched. In the case of maximum power transfer, occurring at equal impedances, mechanical power is equal to the thermal losses as stated by the maximum power transfer theorem for linear networks [11]. This agrees with similar results from Campolo *et al.* showing over 90% dynamic efficiency for a resonant actuator driving a model wing [12].

TABLE 1 Maximal performance for 20 Hz actuation

|  | GM15 | Resonant Actuator |
|---|---|---|
| Torque (N·mm) | 4.04 ± 2.00 | 11.87 ± 0.57 |
| Efficiency (%) | 8.59 ± 0.89 | 96.27 ± 3.20 |
| Pk-Pk Amplitude (°) | 100.0 ± 7.6 | 154.7 ± 4.7 |
| RMS Voltage (V) | 4.53 ± 0.06 | 4.70 ± 0.08 |
| RMS Current (A) | 0.299 ± 0.016 | 0.222 ± 0.006 |

TABLE 2 Maximal performance comparison for 20 Hz actuation

|  | GM15 Motor | GM15 Gearmotor | Metal Gearmotor | Resonant Actuator |
|---|---|---|---|---|
| Weight (g) | 0.9417 | 1.169 | 2.535 | 1.316 |
| Power Density (W/kg) | 263 | 127 | 76 | 974 |
| Torque Density (N·m/kg) | 0.29 | 3.45 | 2.07 | 9.02 |
| Effective Speed (RPM) | 8775 | 351 | 351 | 1031 |

Power density, torque density, and effective operating speed are used to compare the actuators as shown in Table 2. Although the GM15 motor achieved high power density, its coreless inrunner design provides low torque at high speed. A large gear ratio was needed to produce high torque at moderate speeds required for the flapping task. Metal gearboxes are commercially available and offer high efficiency, 78% for a 25:1 ratio [30]. However, their weight is prohibitive and significantly lowers the actuator power density. The 25:1 plastic gearbox of the GM15, despite a lower efficiency of 60% [26, 31], improves power and torque density by 66.8% compared to the metal option. The resonant design increased torque and power density by 161.1% and 666.8%, respectively. The speed is also increased by 193.7%, allowing the FWMAV to generate more lift with larger wing stroke amplitudes. Furthermore, these performance improvements were achieved with over 90% efficiency and 25.8% less current, as stored elastic energy drove acceleration of the load. This significantly reduced heat production, allowing the system to sustain peak output for longer durations. Finally, characterization of the actuator with standard metrics allows researchers to directly determine if it could improve the performance of their system.

### III. POWER DENSITY

Power density is a proposed metric for comparing the performance of actuators in engineered systems as well as to biology, with data compiled in Table 3. Biological muscle is an elastic actuator capable of storing energy, reducing the inertial cost of periodic limb motion. Despite functional similarities, differences are seen between the muscle of the hummingbirds, insects, and cockroaches. On the other hand, the fraction of the body weight devoted to power muscles remains relatively consistent across organism body masses spanning four orders of magnitude. Actuators in robotic systems achieved on average 340% higher power density than their biological counterparts. However, the method of calculating these values produces misleading results. While muscle effectively fulfills the role of a battery, electronics, and actuator, calculations for robotic systems were based on the power actuator mass alone, excluding the weight of transmission elements and steering actuators. It was not possible to rectify these calculations based on published data.

The actuation principles vary amongst robotic systems: Delfly uses a brushless outrunner motor customized to reduce cogging torque, Robobee uses a piezoelectric bimorph, and DASH uses a commercial brushed motor. Our resonant actuator and piezoelectric bimorph produce bi-directional motion to directly drive the wing, although in each case a transmission is still used for torque or displacement amplification. Each wing has its own actuator allowing for differential motion to control the vehicle. Both the Delfly and DASH use large spur gears to increase output torque and include additional transmission mechanisms to convert continuous rotary motor output to limb motion. A single motor drives limb motion, while steering is achieved with additional actuators. Actuator mass fraction varies by system complexity. The Delfly is untethered carrying a battery, necessary electronics, and includes a camera for vision-based autonomy. DASH is also untethered, but lacks sensors for autonomy. Our vehicle and the Robobee are tethered with off-board battery, electronics, and sensors.

TABLE 3 Maximal actuator power densities[1]

|  | Task | Body Mass (g) | Percent Actuator Mass | Power Density (W/kg) |
|---|---|---|---|---|
| **Biological** | | | | |
| Hummingbird [32] *Lampornis clemenciae* | Hover | 8.4 | 29.0 | 309 |
| Hummingbird [32] *Archilochus alexandri* | Hover | 3.0 | 29.0 | 228 |
| Moth [33, 34] *Manduca sexta* | Hover | 1.52[1] | 22.35 | 90 |
| Fruit Fly [35, 36] *Drosophila melanogaster* | Hover | 0.001[1] | 30.0 | 80 |
| Cockroach [37, 38] *Blaberus discoidalis* | Run | 2.6 | 20.4 | 19-25 |
| **Robotic** | | | | |
| **CMU FWMAV** (this study) | **Hover** | **3.22** | **58.49** | **974** |
| Delfly II [5] | Hover | 16.07 | 9.33 | 1000 |
| Harvard Robobee [39, 40] | Hover | 0.08 | 66.67 | 400 |
| Berkeley DASH Hexapod [2] | Run | 16.20 | 19.14[1] | 105 |

[1]Several values in this table were not explicitly published, but calculated to the best of our ability using available data.

Although our actuator and the brushless motor of the Delfly achieve high power density, the bio-inspired resonant actuation principle has several key advantages that reduce vehicle weight and complexity. Our vehicle is fully controlled by differential motion of the two wing actuators, while Delfly uses a power actuator to drive the wing stroke and two steering actuators to control pitch and yaw. Furthermore the crank-rocker transmission that converts motor rotation to reciprocal wing motion increases system weight, complexity, and lowers efficiency compared to the direct drive approach. Due to the lack of elastic elements, energy from wing deceleration cannot be recovered.

In this section we have quantified peak actuator output, which can only be achieved for limited time periods as heating exceeds the thermal dissipation capabilities of the motor. The following sections detail the development of a thermal model that predicts operating limits that allow for sustained hover, while avoiding damage to the actuator.

### IV. THERMAL CIRCUIT MODEL

Heat flow through the actuator was modeled as a second-order lumped parameter thermal circuit shown in Figure 2(a) [27, 28]. The main input was Joule heating, produced at a rate of $I^2R$ in the windings. Bushing friction was identified as a significant secondary source, produced linearly at a rate of $\omega m$ where $\omega$ is the motor speed. The bushing directly contacts the case and shaft, but no significant conductive heat transfer occurs to the windings due to the plastic commutator plate. Therefore bushing friction is treated as a heat input to the case. Each thermal body was modeled as a capacitor and heat flow between bodies as resistors as shown in Figure 2(b). Capacitance was estimated from measured component mass and specific heat capacity for each material. The dominant thermal resistances are convective, expressed as

$$R_{conv} = \frac{1}{hA} \quad (1)$$

where $A$ is the relative surface area and $h$ the heat transfer coefficient based on geometry and fluid flow [27]. Although the convective resistances from the windings to the magnet and the case vary due to differences in geometry, they are difficult to differentiate experimentally. The model combines the magnet and case into a single element, neglecting the conductive resistance between them. The resistance from the windings to this element are then treated as a single resistor. This circular flow travels over the internal cylindrical surfaces in a manner analogous to parallel flow over a flat plate. Therefore, this resistance was modeled as convection to a flat plate with the assumption that flow reaches a steady state velocity profile over each stroke due to the high gear ratio. High rotor velocities are expected to improve air-gap convection by increasing the contribution of forced convection. Forced convection dominates natural convection when the Archimedes number is much smaller than one [27]. At hover, the FWMAV has an effective speed of 841 revolutions per minute (rpm) and an Archimedes number of 1.37e-4.

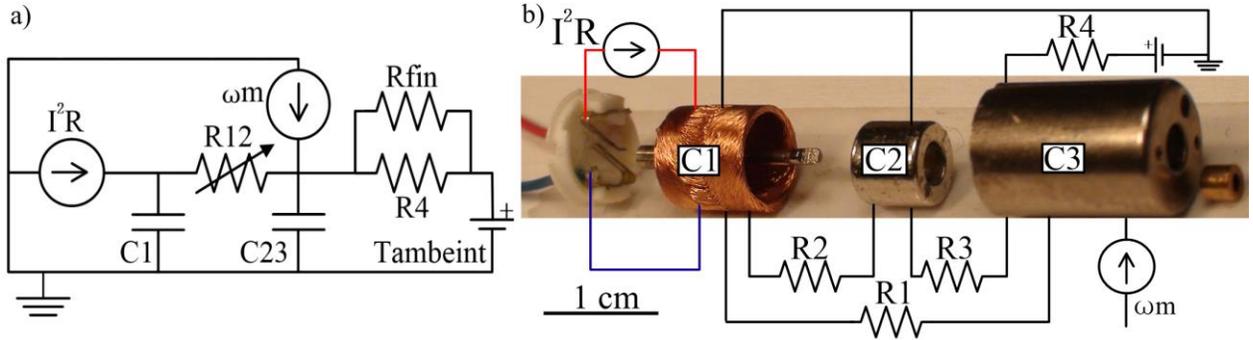

Figure 2. (a) The lumped parameter, thermal circuit model of a DC motor with attached fin used in this study. This reduced order model is derived from the full model shown in b by combining relevant components. (b) Disassembled motor showing brushes, rotor windings, magnet, case, and bushing (left to right). Annotations indicate physical representation of the full thermal model, which includes distinct thermal bodies for each motor component and resistances between all bodies.

Convection from case to atmosphere was conservatively modeled as free convection, representing worst case performance where air around the motor is stagnant. Therefore, a heatsink, or fin, can improve case convection by increasing relative surface area. The heatsink resistance was added in parallel with the convective resistance from the case to atmosphere. This assumes that the heatsink and its connection to the case does not decrease overall flow from the case itself. If the heatsink resistance is too high, the model behaves as if there was no heatsink. Estimated model parameters are listed in Table 4.

TABLE 4 Thermal circuit model parameters

| Symbol | Parameter | Estimate | Tuned |
|---|---|---|---|
| $R_{12}$ | Windings to case & magnet | 201.58 | 33.29 |
| $R_4$ | Case to atmosphere | 622.37 | 154.76 |
| $R_{fin}$ | Heatsink resistance | 9.88 | 467.38 |
| $C_1$ | Winding thermal mass | 0.048 | 0.057 |
| $C_{23}$ | Case & magnet thermal mass | 0.375 | 0.381 |

Thermal resistance has units of Kelvin per Watt (K/W)

## V. MODEL PARAMETER FITTING

The accuracy of lumped parameter models relies on experimental parameter fitting [22]. In order to isolate speed-dependent effects, a dynamometer was created by connecting two motors as shown in Figure 3. The left side, drive motor, was used to spin the right side, sensing motor, without powering it. The case temperature of the sensing motor was measured with a non-contact thermometer and the operating speed measured with a tachometer. The only modification to the motor was a spot of black paint, which increased the emissivity of the steel case for accurate temperature measurements. Previous studies relied on temperatures recorded from internal components for characterization [21]. However, it was not possible to modify the small motor to do so without significantly altering its thermal properties. Following, experimental fitting of parameters for temperature dependent winding resistance, bushing friction, and speed dependent forced convection are discussed.

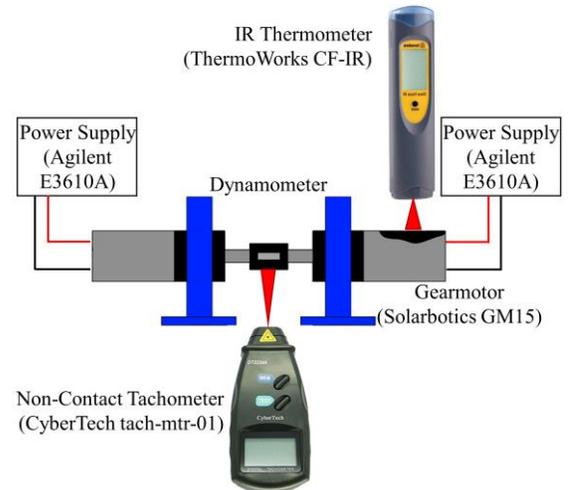

Figure 3. Experimental setup with custom dynamometer, non-contact tachometer, and IR thermometer. Dynamometer experiments allow speed and current dependent heating effects to be isolated.

### A. Temperature Dependent Winding Resistance

Joule heating of the motor windings is expressed as

$$h(t) = I(t)^2 R\big(1 + \alpha(T_w(t) - T_0)\big) \quad (2)$$

where $I(t)$ is the current, $T_w(t)$ is the winding temperature, and R is the winding resistance at $T_0 = 25\,°C$. The temperature coefficient of resistance $\alpha$, is typically assumed from material properties to be 3.93e-3 per degree Celsius for copper [41]. While this is sufficiently accurate for large systems where $I > 1\,A$ and $R < 1\,\Omega$, for our system the resistance term dominates since $I < 0.3\,A$ and $R > 12.5\,\Omega$. To improve accuracy, $\alpha$ was measured experimentally over a relevant range of temperatures, 25 to 155 $°C$, using an oven and determined to be 3.42e-3 per degree Celsius.

### B. Bushing Friction

Friction between the motor shaft and supporting bushing was found to be a significant secondary heat source. This effect was characterized by running the drive motor at constant speeds and measuring the temperature of the sensing motor, which was connected to the power supply in high impedance mode to prevent current from flowing due to the back electromotive force. Experimental data is shown in Figure 4. The relationship between steady state temperature and speed was linear ($R^2 = 0.99$) over the tested range. This effect was added to the circuit model as a heat source at the input of C23 proportional to speed with slope *m*. The steady state temperature depends only on the value of *m* and *R4*. A linear expression for *m* as a function of *R4* was experimentally determined using estimated values and used in the final parameter fitting.

### C. Speed-Dependent Convective Thermal Resistance

At hover, the Archimedes number indicates significant forced convection. However, its effect on overall heat flow and dependence on speed needed to be characterized. For this, a pulse-spin experiment was used. First the sensing motor was stalled at a current of 570 mA for 5 s. This pulse was less than one-tenth of the system time constant and elevated case temperature to 70 °C. The drive motor was then run at a constant speed. The temperature of the sensing motor was recorded as it cooled to steady state, which varied with speed due to bushing friction. The settling time, which depends on the internal convective resistances *R12*, was calculated as shown in Figure 5. Limitations of the dynamometer made it difficult to conduct precise replicate trials. Individual experiments are denoted by marker type. Each experiment was fit with a linear trend and the average of these was used to create the aggregate fit line shown in black. The strong trend of decreasing settling time indicates that *R12* decreases linearly with increasing speed.

The circuit was implemented in Matlab Simulink using SimPowerSystems Specialized Technology library. Model parameters were trained with genetic optimization (Matlab ga) using the squared difference between case temperature from pulse-spin experiments and model as the fitness value. The optimization was initialized with the parameter estimates shown in Table 4. The bounds on capacitance values were set to ±20% to account for variation in material properties. The resistance terms were allowed to vary from 1 up to the estimated value, since convective resistance tends to be over-estimated. Optimization was performed over three sets of pulse-spin data at 0, 490, and 1363 rpm. R4, C1, and C23 showed close agreement between datasets and an averaged value was used for the final simulation. As predicted *R12* was seen to vary linearly with speed and was fit as follows,

$$R12 = 33.29 - 0.034\omega. \quad (3)$$

Results for the trained model are shown in Figure 6.

Training of the simulation resulted in the tuned parameters listed in Table 4. *C23* is closely estimated since the case and magnet are both single materials. *C1* was underestimated as the windings were assumed to be pure copper, but contain plastic wire coating and binding epoxy.

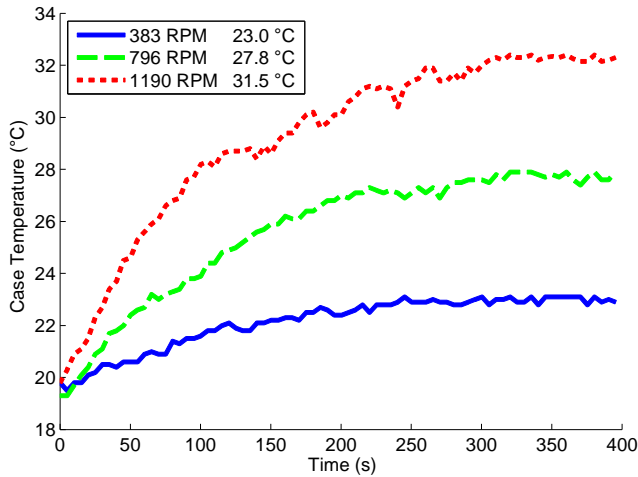

Figure 4. Case temperature vs. time with speeds indicated in the legend.

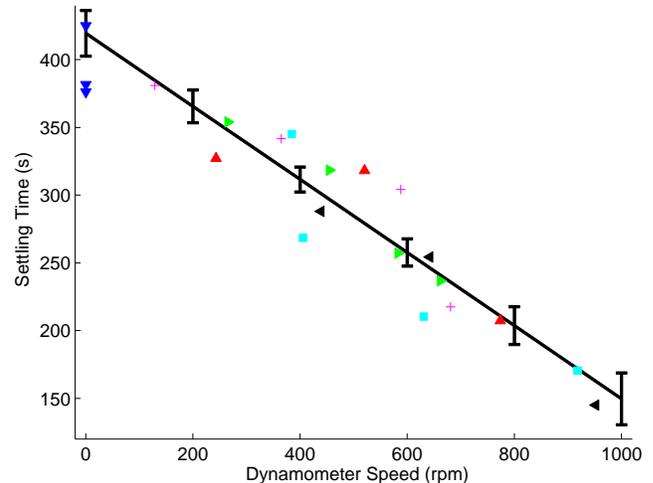

Figure 5. Pulse-spin experiment settling time vs. operating speed results

Convective resistances *R12* and *R4* were significantly overestimated. *R12* was difficult to estimate due to the unknown nature of the motor's internal flow. The decrease in *R4* is likely due to the air around the motor not being fully stagnant. For training of the heatsink data, *C23* was augmented by the heatsink mass, approximately doubling its value. *Rfin* was significantly higher than predicted indicating low conductive flow through the thermal adhesive used to attach it to the motor and also suggests that *R4* was increased by the heatsink addition.

The trained simulation accurately predicts case temperature as shown in Figure 7. The actuator is tested to a safe limit at the hover operating point, which corresponds to a current of 0.24 A and a speed of 841 rpm. It takes the system 41 s to reach critical temperature and 352 s to return to within two degrees of room temperature. The case temperature shows very close agreement, particularly during heating. A slight temperature offset can be seen during the midway through the cooling phase, however, the settling time is still captured correctly. The winding are predicted to have a slightly higher temperature than the case during heating, with a difference of 6 °C at the peak. During cooling, the windings quickly return to the case temperature due to the low thermal resistance and small thermal mass relative to the case. A case temperature upper bound was used for the safe operating envelope of the actuator. The case should not exceed 80 °C as this deforms and eventually melts the plastic gearbox. In practice, the windings should also not exceed 105 °C as this significantly degrades the wire's insulating coating leading to permanent failure upon melting [42]. However, the simulation indicates that the case temperature limit also protects the windings.

In this section we have outlined a novel thermal model to accurately predict heating dynamics of a micro-motor. The critical effects of variable winding resistance, bushing friction, and speed-dependent forced convection were characterized experimentally. Furthermore, all model training data was collected with the motor alone operating continuously, while validation was performed with the resonant actuator undergoing reciprocal operation. The trained model accurately predicts case temperature for the resonant actuator at the hover operating point and sets a safe operating limit. Winding temperature is also predicted and indicates that a case temperature limit also protects the windings from thermal damage.

## VI. HEATSINK DESIGN

The circuit model was used to determine maximum total heatsink resistance that improved performance. For hover, corresponding to a current of 0.24 A, preliminary simulations indicated the threshold to be 10 K/W. A design optimization was then performed to minimize the weight as follows,

$$\min_{segment\ \#} \left( \min_{\boldsymbol{\theta}} (Weight_{array}) \right) \quad (4)$$
$$s.t.: Biot\ number\ of\ each\ segment < 0.1$$
$$Heatsink\ resistance \leq 10\ K/W$$

The heatsink was modeled as an array of fins composed of as many connected segments, with heat transferred between them by conduction and to the atmosphere by convection. The Biot number constraint was needed to ensure a physically accurate solution where the heat being drawn out of each segment never exceeded that heat transferred in. Initial simulations focused on several long fins that took advantage of the high velocity downwash from the wings. However, an array of many shorter fins was determined to be weight-optimal. A total resistance of 9.88 K/W was predicted for an array of 42 copper fins, where each fin was 25 mm x 0.1 mm x 2 mm. The design was laser cut (LPKF ProtoLaser U3) from a copper sheet and attached to the motor with a thermal adhesive (Arctic Alumina AATA-5G) as shown in Figure 8. The total weight was 0.97 g per motor, which is within the payload of the vehicle. Given that this design relies on free convection alone, it could be generalized to other applications. With the heatsink the system takes 83 s to reach critical temperature and 260 s to return to within two degrees or room temperature. The fin increases system operating time by 102.4% and reduces cooling time by 26.1%.

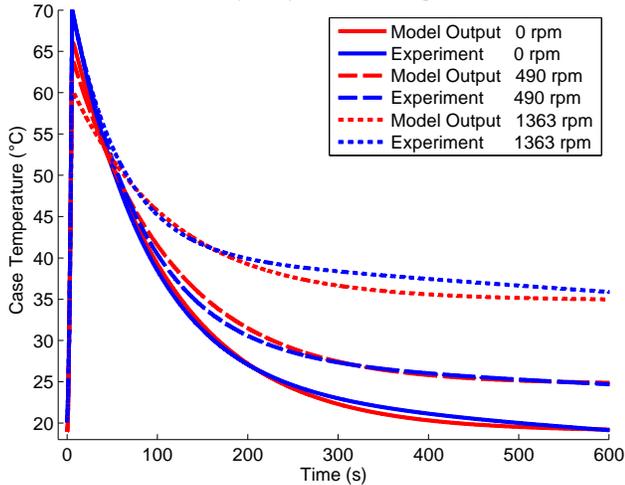

Figure 6. Results of the pulse-spin experimental data from the motor only in blue, compared to the output of the trained simulation in red.

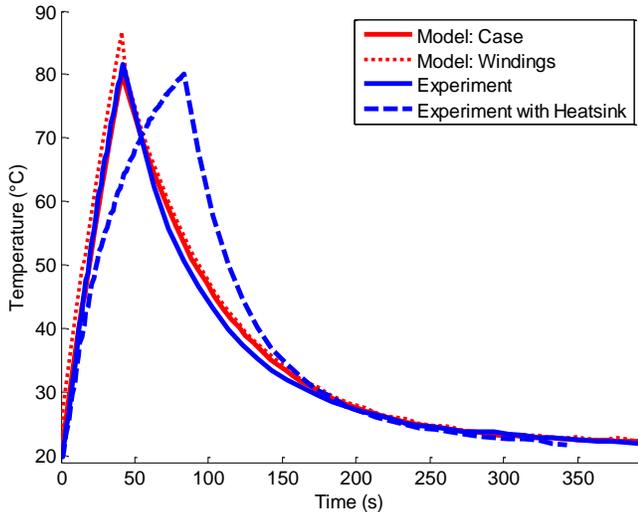

Figure 7. Comparison of experimental and model data for the resonant actuator operating at hover (0.24 A, 841 rpm) showing both core and case temperatures. Case temperature for the actuator with heatsink is also shown.

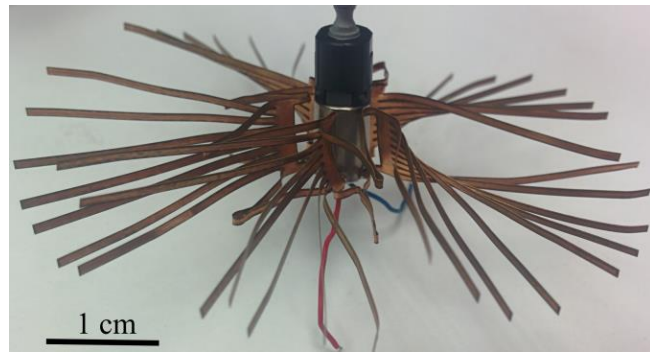

Figure 8. Fabricated weight-optimal heatsink attached to motor.

## VII. CONCLUSION

Resonant actuation serves as a general principle by which oscillating limbs can be efficiently driven by motors in miniature robotic systems. Furthermore, directly driving independent limb motion can eliminate the need for complex transmissions and simplify control. Compared to the gearmotor alone, the resonant design is shown to increases torque and power density by 161.1% and 666.8% respectively, while decreasing the drawn current by 25.8%. Characterization with standard metrics allows researchers to determine if such an actuator could be incorporated into their work. Measured actuator efficiency exceeds 87% for amplitudes above 32.75°. Increasing efficiency, resulting in decreased current draw, lengthens the time that peak performance can be sustained. However, for applications where continuous high torque is required, these improvements may not be sufficient for extended operation.

To determine a safe operating envelope for the system, a thermal modal of the actuator was developed. A lumped parameter thermal circuit was experimentally fit using current and speed as inputs and case temperature as the output. In order to accurately model the micro-motor, two orders of magnitude smaller than those previously characterized, the effects of temperature-dependent winding resistance, bushing friction, and speed-dependent forced convection were experimentally determined using a custom dynamometer. The trained model accurately predicts the time course of case temperature for hovering of the FWMAV and subsequent cooling. A safe operating envelope for the actuator is determined to prevent permanent damage. Furthermore, the model was used to design a weight-optimal heatsink that relies on free convection and increases system operating time by 102.4%. Resonant actuation and thermal modeling are powerful tools to better understand and maximize the performance of electromagnetic actuators in miniature mobile robots and could be considered by researchers for a variety of applications.

## VIII. ACKNOWLEDGEMENTS

The authors would like to thank Dr. Aaron Johnson, CMU Robomechanics Lab, for his support during this project and advice on experimental setup. We would like to thank Jesse Roll, Purdue Biorobotics Lab, for providing feedback on the manuscript.

*Technologies for Practical Robot Applications (TePRA), 2011 IEEE Conference on*, 2011, pp. 64-69.